\documentclass[aps,prd,nofootinbib,preprint,tightenlines,showpacs]{revtex4-1}
\newcommand{\be}{\begin{equation}}
\newcommand{\ee}{\end{equation}}
\newcommand{\bea}{\begin{eqnarray}}
\newcommand{\eea}{\end{eqnarray}}
\usepackage{graphicx}

\def\bx{\mathbf{x}}
\def\ba{\mathbf{a}}
\def\bb{\mathbf{b}}

\def\coloronline{}

\begin{document}

\title{Cosmic string loop shapes}

\author{Jose J. Blanco-Pillado${}^{a,b}$}
\email{josejuan.blanco@ehu.es}
\author{Ken D. Olum${}^{c}$}
\email{kdo@cosmos.phy.tufts.edu}
\author{Benjamin Shlaer${}^{c}$}
\email{shlaer@cosmos.phy.tufts.edu}
\affiliation{$^a$ Department of Theoretical Physics, University of the Basque Country,  Bilbao, Spain}
\affiliation{$^b$ IKERBASQUE, Basque Foundation for Science, 48011, Bilbao, Spain}
\affiliation{$^c$ Institute of Cosmology, Department of Physics and Astronomy,\\ 
Tufts University, Medford, MA 02155, USA}

\begin{abstract}

We analyze the shapes of cosmic string loops found in large-scale
simulations of an expanding-universe string network.  The simulation
does not include gravitational back reaction, but we model that
process by smoothing the loop using Lorentzian convolution.  We find
that loops at formation consist of generally straight segments
separated by kinks.  We do not see cusps or any cusp-like structure at
the scale of the entire loop, although we do see very small regions of
string that move with large Lorentz boosts.  However, smoothing of the
string almost always introduces two cusps on each loop.  The smoothing
process does not lead to any significant fragmentation of loops that
were in non-self-intersecting trajectories before smoothing.

\end{abstract}


\maketitle

\section{Introduction}

Cosmic strings are microscopically thin, astrophysically long objects
which may have formed at early times as a result of phase transitions
or brane inflation in superstring theories.  See Ref.~\cite{Alexbook}
for a review.  In the simple cases to be discussed here, cosmic
strings have no vertices or ends, and thus form a ``network''
consisting only of infinite strings and closed loops.

When cosmic strings intersect each other, they can reconnect.  In
the case of strings formed by phase transitions in field theory, this
reconnection nearly always takes place.  But in the case of
strings from superstring theory, the strings can pass through each
other unchanged, only reconnect with some probability $p$ in the range
$10^{-3}$ to 1.  When a string reconnects with itself, the result is to
emit some portion of the string into a closed loop.  Closed loops
oscillate, emit gravitational waves, and so eventually decay away.

This process allows the long string network to scale with the
expansion of the universe, meaning that any statistical measure of the
string network properties with dimension (length)$^\alpha$ goes as
$t^\alpha$, where $t$ is the the age of the universe.  In particular,
the density of long strings (length per unit volume) goes as $t^{-2}$,
and thus scales as radiation in the radiation era and as matter in the
matter era.  Scaling means that the energy density of strings is a
fixed fraction of the total energy density in the radiation and matter
eras.  Thus they do not dominate the universe as monopoles
would  \cite{Zeldovich:1978wj,Preskill:1979zi}, nor do their effects become ever tinier with time in these
eras.

The evolution of a cosmic string in flat space is easily described.
Let $\bx(\sigma,t)$ describe the spatial position of the string at time $t$,
and let the parameter $\sigma$ be chosen to parameterize the string
energy in units of $\mu$, the energy per unit length of a static
string.  Then one can show that in this gauge, 
$\bx'^2 + \dot\bx^2 = 1$, where prime and dot denote
differentiation with respect to $\sigma$ and $t$ respectively.  By
appropriate choice of starting point for $\sigma$ at different $t$, we
can make $\bx' \cdot \dot\bx = 0$.  The equations of motion then
become
\be\label{eqn:eom}
\ddot\bx(\sigma, t) = \bx''(\sigma, t)\,.
\ee
The general solution can be written
\be\label{eqn:xab}
\bx(\sigma, t) = {1\over 2}\left[\ba(t-\sigma)+\bb(t+\sigma)\right]\,,
\ee
where $\ba$ and $\bb$ are any functions obeying the constraint $|\ba'|
= |\bb'| = 1$.  In the expanding universe, the solutions are more
complicated, but Eqs.~(\ref{eqn:eom},\ref{eqn:xab}) are good
approximations for loops much smaller than the Hubble size.

Because the functions $\ba'$ and $\bb'$ have unit magnitude, they trace
out paths on the unit sphere, the so-called Kibble-Turok (KT) sphere.  In general, one would expect those
paths to intersect \cite{Kibble:1982cb,Turok:1984cn,Vachaspati:1987rw}.  In particular,
in the case of a closed loop observed in its rest frame, the center of
gravity of $\ba'(\sigma)$ and $\bb'(\sigma)$ lies at the center of the
sphere, and so these paths will intersect except in special cases.

At an intersection, where $\ba'(t-\sigma) =\bb'(t+\sigma)$,
Eq.~(\ref{eqn:xab}) gives $\bx' = 0$ and $\dot\bx = 1$.  Thus the
string doubles back on itself at a point which (in the approximation
of an infinitely thin string) moves at the speed of light.  Such a
point is called a cusp \cite{Turok:1984cn}.  At a cusp, the string
core may overlap with itself, leading to the emission of high-energy
particles
\cite{Brandenberger:1986vj,Mohazzab:1993ah,BlancoPillado:1998bv,Olum:1998ag}.
The high Lorentz boost near the cusp may also produce beams of
gravitational \cite{Vachaspati:1984gt,Damour:2000wa,Damour:2001bk} or
electromagnetic radiation
\cite{Vilenkin:1986zz,Spergel:1986uu,BlancoPillado:2000xy,Cai:2011bi},
and if the string is coupled to other fields it may lead to particle
production
\cite{Srednicki:1986xg,Damour:1996pv,Peloso:2002rx,Sabancilar:2009sq,Long:2014mxa,Long:2014lxa}.
Many suggestions have been made for observable astrophysical
signatures coming from cosmic string cusps. It is therefore important
to understand the parameters that characterize these kinds of events
in a typical cosmic string loop produced in a scaling network of
strings.

The above argument applies only if strings are smooth.  The usual
calculations, in fact, assume that $\ba''$ and $\bb''$ are of order
$1/L$, where $L$ is the invariant length of the string loop.  On the other 
hand, cosmic strings have kinks: places where the direction of the string changes suddenly
due to a previous reconnection.  Such discontinuities in $\ba'$ and
$\bb'$ allow them to jump to different parts of the unit sphere
without crossing each other.  In such cases, the argument fails, and
there may be no cusps.  To know whether or not there are cusps, we need
to know the shape of the functions $\ba$ and $\bb$.

A loop is born when a long string reconnects with itself.  At this
time, the loop consists of structures that were present on the long
string from which it formed, plus two kinks from the final
reconnection.  In the first oscillation, the loop may fragment, but
after one oscillation fragmentation is rare.  (In flat space, loop
trajectories are periodic, so after one whole oscillation without
fragmentation, no further fragmentation is possible.  In the expanding
universe there is a correction, but it is typically small.) After
that, the loop slowly loses energy by emitting gravitational waves,
with power $\Gamma G \mu^2$, where $\Gamma$ is a number 
of order 50 \cite{Vachaspati:1984gt,Turok:1984cn,Burden:1985md,Garfinkle:1987yw}
and $G$ is Newton's constant.  This leads the loop to lose length at
rate $\Gamma G \mu$.  When it has lost a significant fraction of its
length by this process, its shape may be substantially modified.  In
particular, we expect it to be much smoother than it was at formation.

The loops which are most important for observable signals are those
which have significantly evaporated.  At any given time $t$ in the
history of the universe, most string is in loops of size around
$\Gamma G \mu t$, which are thus roughly half evaporated
\cite{Blanco-Pillado:2013qja}.  Even if we consider loops of some
fixed size $\ell$, most of these were created long ago
\cite{Blanco-Pillado:2013qja} and have suffered significant
evaporation, except for those of the very largest sizes.

The goal of this paper is to determine the shape of cosmic string loops
over the course of their lives.  We start by analyzing loops and long
strings taken directly from a large cosmic string network simulation
in the expanding universe.  We find that these strings consist
primarily of long segments that are generally straight, connected by
large-angle kinks.  The straight segments are not exactly straight,
but rather consist of several shorter and straighter segments, with
smaller-angle kinks between those.  We never see a smooth motion of
$\ba'$ or $\bb'$ from one place to another of the sort that could lead
to the formation of a cusp.

Thus the track of $\ba'$ or $\bb'$ in the simulated loops at formation
consists not of a smooth, closed curve, as usually envisioned, but
rather of a sequence of small, irregular regions.  Within each region,
the tangent vector jumps rapidly from place to place, and the regions
are connected by sudden, large jumps.  In a loop, these regions are
either well separated or touch along a curve, but they never
significantly overlap or cross through each other.  Thus the
``crossings'' of $\ba'$ and $\bb'$ occur entirely in the large jumps
(kinks) and never in smooth regions of string that would lead to the
phenomenology associated with cusps.

Our large scale string network simulation does not include gravitational back reaction, so
it is accurate only at scales above those that might be smoothed by
that process.  But gravitational effects on long strings, and thus on
newly formed loops, are not very important.  Loops forming at any
given time have sizes much larger than $\Gamma G \mu t$ \cite{BlancoPillado:2011dq} so their
general shapes are little affected by previous smoothing.

Gravitational effects after loop formation, however, are very
important, as discussed above.  Unfortunately, we are not presently
able to simulate the gravitational back-reaction on a loop.  So
instead, we attempt to understand it by smoothing the loop using
convolution, as described below.  This gives us a more realistic
population of loops with various degrees of smoothing, and we study
these in addition to the newborn loops.  Indeed, after significant
smoothing, most loops have two significant cusps in each oscillation.

The rest of this paper is organized as follows. In the next section,
we discuss the shape of loops directly taken from the simulation
without gravitational smoothing, and in Sec.~\ref{sec:long} we discuss
the structures found on long strings.  In Sec.~\ref{sec:smoothing} we
discuss our smoothing procedure, and in
Sec.~\ref{sec:smoothingresults} we discuss the shape of loops with
various degrees of smoothing, the number and size of cusps they
produce, their distribution of angular momenta, and to what degree the
trajectory of each loop lies on a plane.  We conclude in
Sec.~\ref{sec:conclusion}.

\section{Loops at formation}\label{sec:formation}

We first discuss the shape of cosmic string loops just as they are
found in our simulation, which does not included gravitational
smoothing.  Our simulation process is described in detail in
Refs.~\cite{BlancoPillado:2011dq,BlancoPillado:2010sy}.  The results 
in this section are taken from a simulation in the matter era of 
size 500 in units of the initial correlation length. The initial 
conformal time for this simulation is 4.5 and the ending time 
500.0, so the dynamic range is 110.

Several groups have studied the properties of non-intersecting
loops from a set of random initial conditions \cite{Scherrer:1990pj,Copi:2010jw}. We
will make some comparisons with these other studies when appropriate. On the other 
hand, our results should describe more accurately the properties of real loops from
a network since we take the initial distribution of loops from 
a scaling network of strings. In fact we believe these loops
are a good representative set of scaling loops themselves.  (See the
discussion in Ref.~\cite{BlancoPillado:2011dq}).

Because $\ba'$ and $\bb'$ are piecewise linear in our simulation,
their paths on the unit sphere consist of discrete points
corresponding to the linear pieces.  Where the string changes
direction at a kink, there is a line connecting two points.

\begin{figure}
\includegraphics[width=6in]{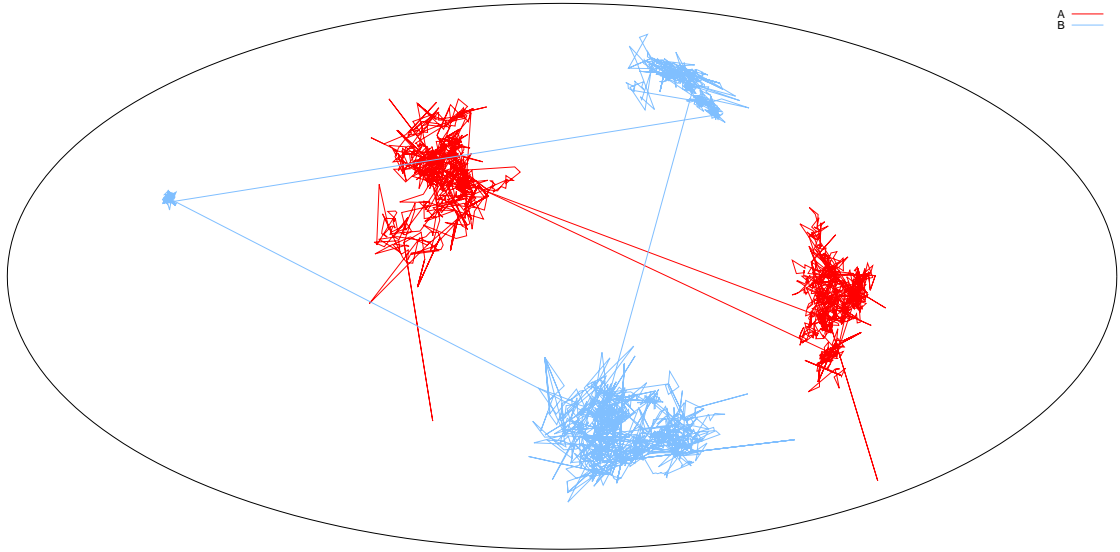}
\caption{\coloronline The paths of $\ba'$ (red or dark) and $\bb'$
  (blue or light) on the unit sphere for a loop of length 6.28 seen at
  time 500.0.  Each vertex represents a segment with constant tangent
  vector.  The tangent vectors have been projected from the unit sphere
  into the figure plane with the Mollweide projection (the same one
  usually used for cosmic microwave background maps).  Kinks are
  represented by straight lines on the figure.}
\label{fig:separated}
\end{figure}

\begin{figure}
\includegraphics[width=2.5in]{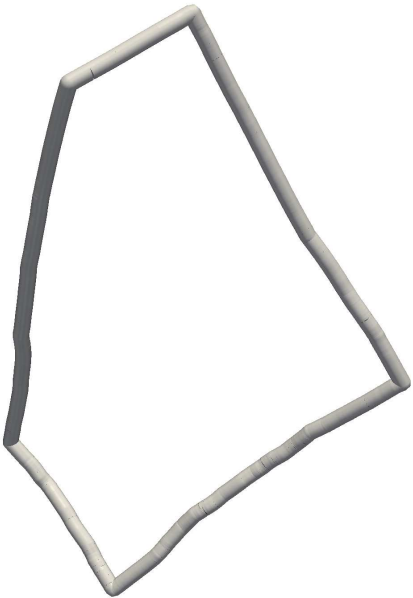}
\caption{A snapshot of the loop represented by its $\ba'$ and $\bb'$
  in Fig.~\ref{fig:separated}. One can clearly see the 5 different
  orientations of the tangent vector of the string corresponding
  the 5 combinations of the regions of $\ba'$ and $\bb'$ in the
  Kibble-Turok sphere in Fig.~\protect\ref{fig:separated} that exist
  at this time. The small variations of the tangent vector in each of
  these segments give some idea of the angular spread of the lumps on
  the KT sphere.}
\label{fig:5-kink-loop}
\end{figure}

In Fig.~\ref{fig:separated}
we show the paths of $\ba'$ and $\bb'$ for a non-self-intersecting
loop appearing in our simulation at time 500.  We see immediately that
the tangent vectors are arranged in five separated clumps, with jumps
between them.  In this particular case, the clumps of $\ba'$ are well
separated from the clumps of $\bb'$.  The only places where the paths
of $\ba'$ and $\bb'$ cross each other are in large-angle kinks that
connect the different clumps.  Thus this loop will not produce any of
the traditional cusps. We show in Fig.~\ref{fig:5-kink-loop} a
snapshot of this loop in physical space at a time when
one can clearly see the five different directions made from these
blobs.

\begin{figure}
\includegraphics[width=6in]{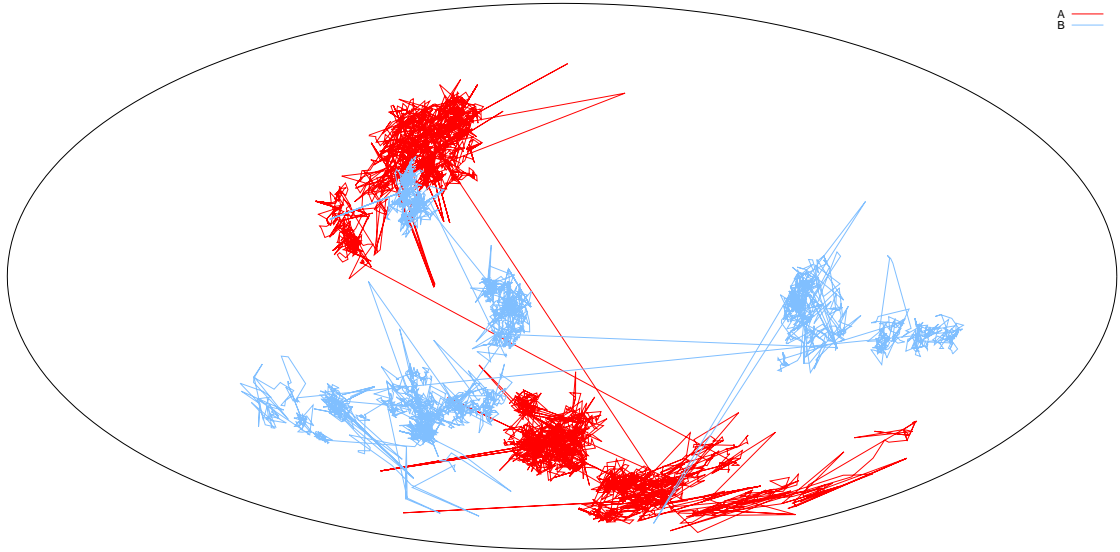}
\caption{\coloronline The paths of $\ba'$ (red or dark) and $\bb'$
  (blue or light) on the unit sphere for a loop of length 8.29.}
\label{fig:join}
\end{figure}

\begin{figure}
\includegraphics[width=5.7in]{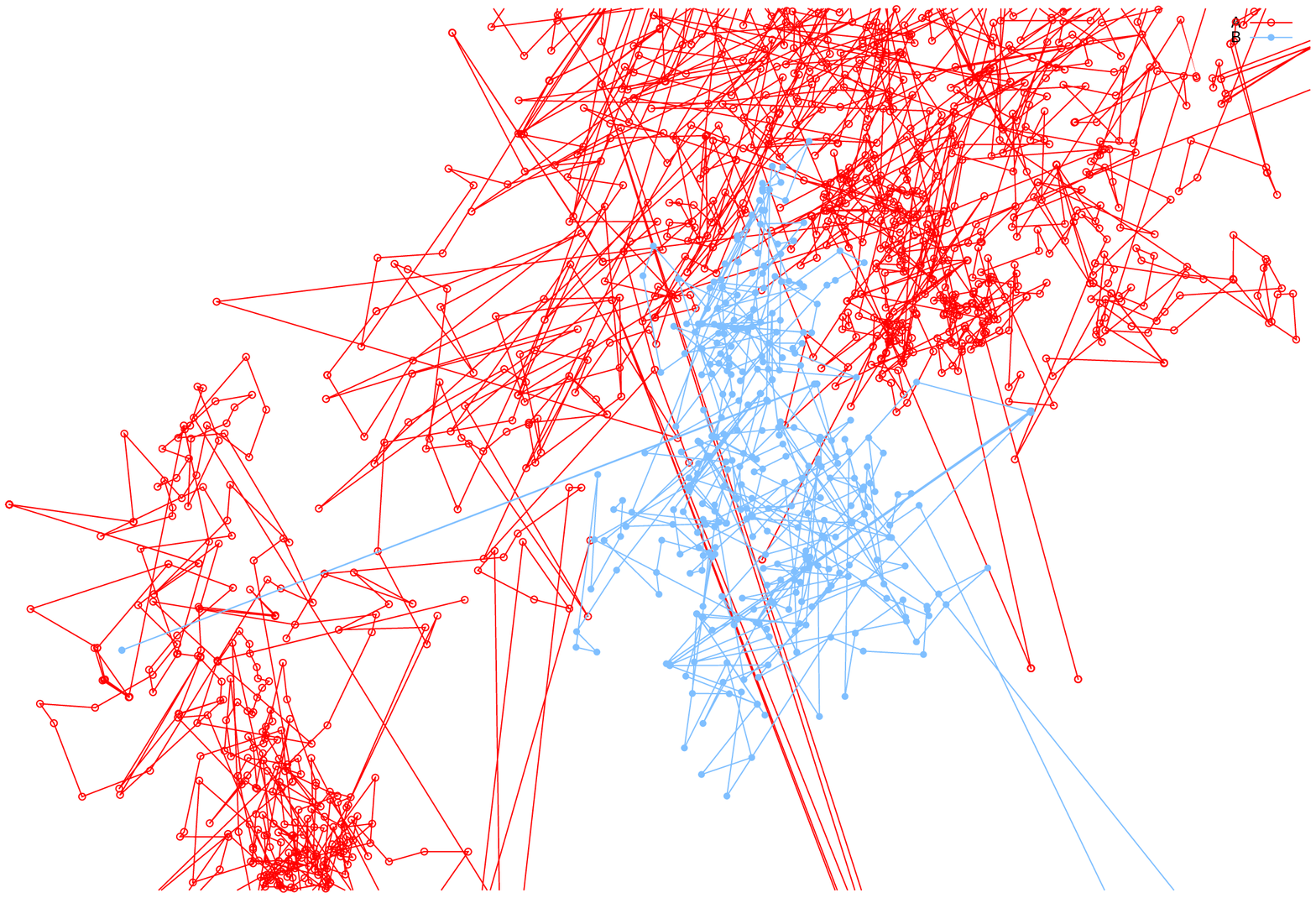}
\caption{\coloronline Close-up of the upper left of
  Fig.~\protect\ref{fig:join}.}
\label{fig:closeup}
\end{figure}

A more common situation is shown in Fig.~\ref{fig:join}.
In this case the clumps of $\ba'$ and $\bb'$ touch along a curve,
further shown in Fig.~\ref{fig:closeup}.
This happens because overlapping segments are lost to loop production,
until the unit sphere has been divided into domains with many points
of $\ba'$ and those with many $\bb'$, separated by a complicated
curve. This is the telltale sign of a non-self-intersecting loop on
the Kibble-Turok sphere. The non-self-intersecting loops resulting
from our simulations cover a continuum of possible distributions of
clumps on the sphere, from the kind of well separated clumps in
Fig.~\ref{fig:separated} through the type represented by the first
panel in Fig.~\ref{fig:smoothab} below, where most of the sphere is
covered by the blobs.

One can disturb this situation by introducing additional perturbations
to the loops so that the ``islands'' of $\ba'$ heavily overlap the
ones of $\bb'$. We have performed such experiments with a number of
loops. The result from this exercise is an excited loop whose
subsequent evolution quickly produces an important amount of
fragmentation until one is back again in a similar representation of
the loop on the KT sphere of isolated blobs of $\ba'$ and $\bb'$
segments.

Near the dividing curve between the $\ba'$ and $\bb'$ clumps, there will be points of $\ba'$ and points of
$\bb'$ that are very close to each other.  In fact, there are even
some points that lie on the ``wrong side'' of this curve, i.e., places
where points of $\ba'$ reach into the domain of $\bb'$, and vice
versa.  In most such cases, the distance (angle) on the wrong side of
the curve is quite small, but there are a few cases, such as the
leftmost point of $\ba'$ in Fig.~\ref{fig:closeup}, where $\ba'$
reaches deeply into the domain of $\bb'$.  Such deep excursions seem
to consist only of single points with only a very tiny amount of
string.  For example, the point mentioned above represents a segment
of $\ba'$ with length less than $10^{-8}$.

When $\ba'$ and $\bb'$ are nearly equal, a piece of string will move
rapidly.  This may lead to gravitational wave emission and other
signatures.  In some ways, these regions are similar to cusps.
Indeed, they may be an important source of gravitational waves, which
will be the subject of a separate paper.  However, they do not have
the usual form of the cusp where $\ba'$ and $\bb'$ move steadily
across the sphere and cross each other.  They certainly cannot be
analyzed by Taylor series expansion around a crossing point, as is
usually done for cusps.

\section{Long strings}\label{sec:long}

\begin{figure}
\includegraphics[width=6in]{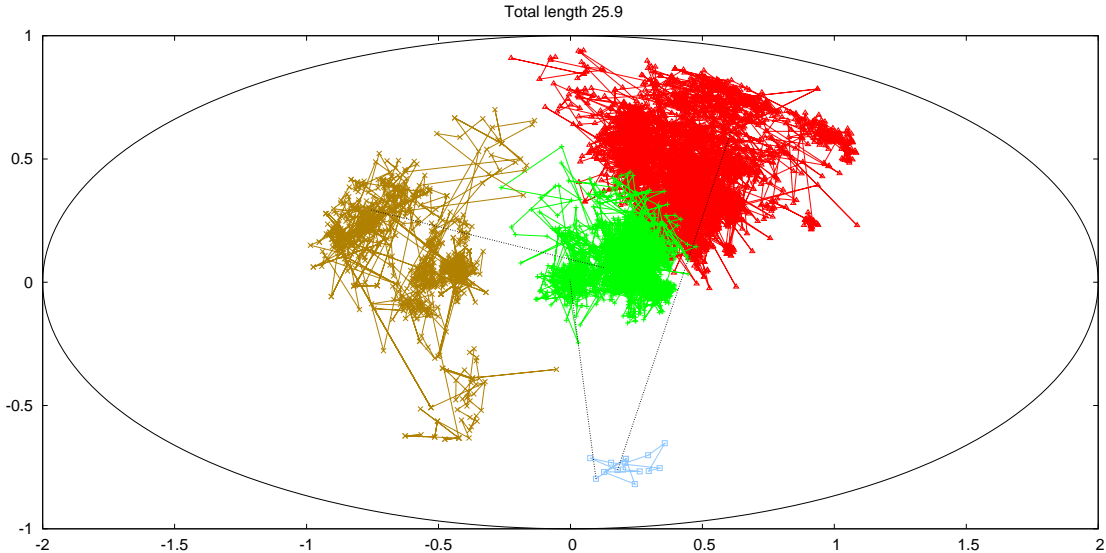}
\caption{\coloronline 13197 consecutive segments of a long string at
  time 500.0, separated into 4 clumps by a hierarchical clustering
  algorithm.  The clumps are shown with different colors (shading) and
  symbols.  Dotted lines show the kinks that connect the clumps.}
\label{fig:long}
\end{figure}

We can gain some further insight into the shape of strings by looking
at the tangent vectors to the long strings in our simulation.  In
Fig.~\ref{fig:long} we show a section of $\ba'$ of length 25.9
consisting of 13197 consecutive straight segments of
string\footnote{The number 13197 was chosen to start and end at large
  kinks.}.
The colors are chosen by a hierarchical clustering algorithm, which
works as follows.  We first merge the two segments of $\ba'$ which are
the closest on the unit sphere, replacing them by a merged segment with
the total length and the average direction.  The average is weighted
by the length of each segment.  We then proceed to the next-closest
pair, and so on.  When the angle between two pieces of string (which
may be the results of previous merges) is greater than a threshold,
here $60^\circ$, we stop merging.

The clustering procedure has divided the path of $\ba'$ into 4
``clumps''.  Within each clump, the points appear to wander over a
certain region of the unit sphere.  Large kinks connect each clump to
the next.

\begin{figure}
\includegraphics[width=4in]{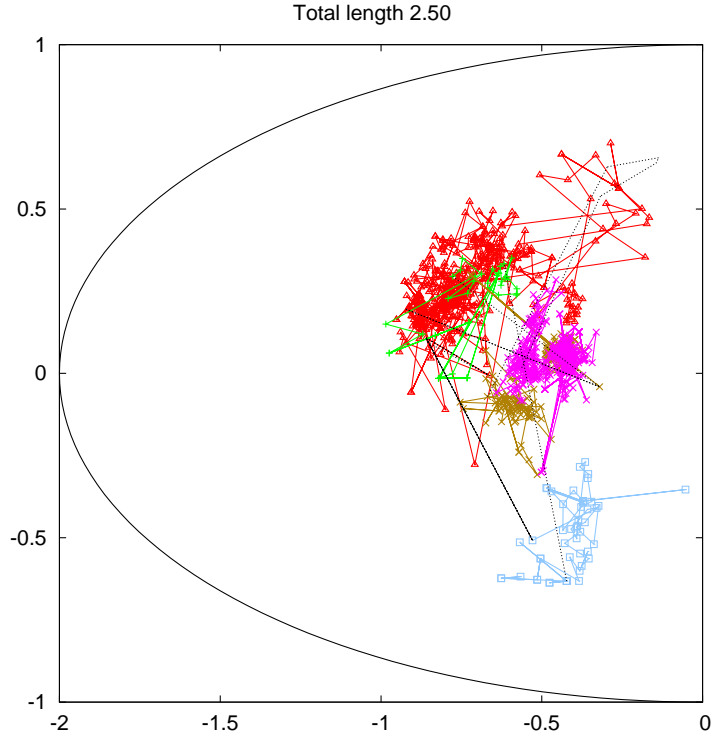}
\caption{\coloronline The leftmost clump of
  Fig.~\protect\ref{fig:long}, further broken down with threshold $40^\circ$}
\label{fig:long-2}
\end{figure}

Figure~\ref{fig:long-2}
shows only the leftmost clump from Fig.~\ref{fig:long}, further split
by the same algorithm with the threshold now $40^\circ$.  The motion
within the clump is not just random wandering.  In fact the clump
can be seen to be made up of 5 major sub-clumps with a certain
degree of overlap.  Such a structure is what one would expect in a
scaling regime.  If one separates clumps at a certain angular
threshold, the size of the clumps as a fraction of the horizon size
should not change with time.  Thus clumps at later times should be
composed of many clumps that existed at earlier times.  These clumps
are made smaller by damping due to the expansion of the universe.
Thus at any given time, each large clump should consist of several
smaller clumps separated by smaller angles, and so on.

\begin{figure}
\includegraphics[width=3in]{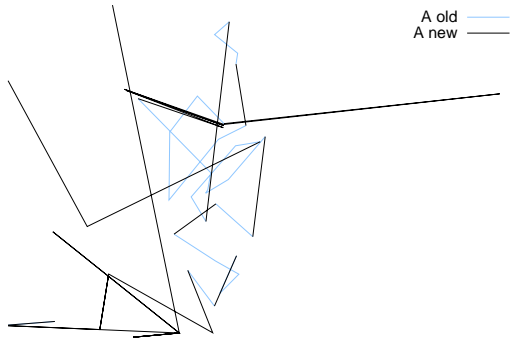}
\caption{\coloronline Magnified view of the bottom clump of
  Fig.~\protect\ref{fig:long}.  Kinks that were present in the initial
  conditions are shown in blue (gray); those that were formed by later
  intersections are shown in black.  The latter includes the kinks
  that go into and come out of the clump.  The path of $\ba'$ sometimes
  jumps to a new value and then back again, leading to lines that
  appear to end.}
\label{fig:creation}
\end{figure}

Since our simulation begins with piecewise-linear initial conditions,
one might worry that the clumping structure is related to the kinks
that were present in the initial conditions.  To investigate this
possibility, in Fig.~\ref{fig:creation}
we distinguish the kinks that were present in the initial
conditions from those occurred later at reconnections.  While some of
the structure comes from the initial conditions, a significant portion
was added later.

Our picture of string shapes is thus the following.  The tangent
vector $\ba'$ on a long string wanders over small areas of the unit
sphere to form small clumps, then jumps to another nearby area to form
another small clump.  Nearby small clumps form a single larger clump,
and larger jumps connect the larger clumps, and so on.  The path of $\ba'$ or $\bb'$ on a loop at formation is simply a
section of the path that existed in a long string, with the ends of
that section connected by one new kink formed when the loop was broken
off, thus it has the same structure, and we do not expect cusps on
newly-formed loops.

\section{Gravitational smoothing}\label{sec:smoothing}

Now we turn to the effects of gravitational back reaction on the shape
of strings.  The general expectation, going back to the work of
Quashnock and Spergel \cite{Quashnock:1990wv} is that kinks formed at
time $t_1$ will be smoothed out by a later time $t_2$ over length
$\Gamma G\mu (t_2 - t_1)$, where $\Gamma\sim 50$.  This effect limits
the depth of the hierarchy of clumps and sub-clumps.  Since the kinks
in small sub-clumps were generally formed long ago, sub-clumps with
$\sigma < \Gamma G\mu t$ do not exist at time $t$.  Instead, at such
scales the string is smooth.

On long strings, or recently formed loops, this effect is minimal.
Observational constraints limit $G\mu$ to be less than $10^{-8}$
(E.g., see Ref.~\cite{Blanco-Pillado:2013qja}) so $\Gamma G\mu <
10^{-6}$.  Only a tiny percentage of the energy goes into loops with
sizes smaller than or comparable to $10^{-6} t$, so in general
the effects of smoothing are small.

On the other hand, for old loops, the effect is very different.  For a
loop which has lost half its initial energy to gravitational
radiation, the length of the loop and the smoothing scale are equal.
Such a loop will be quite smooth and will likely have cusps at scales
similar to its length.  As discussed in the Introduction, loops with a
significant degree of evaporation are the most important
cosmologically, so it is crucial to understand the effect of
gravitational back reaction on loop shapes.

Unfortunately, we do not know the effect of gravitational back
reaction in detail, so we must resort to a toy model to understand
these phenomena.  As a loop radiates gravitational waves, it must lose
energy.  We expect small structures to radiate their energy away more
quickly than large ones, so that small structures are damped more
quickly, resulting in a smoothing of the loop.  Thus we will attempt
to understand effects of gravitational back reaction by progressively
smoothing our loops using convolution.

There is a simple model of gravitational back-reaction for wiggles on
a straight string \cite{Hindmarsh:1990xi,Siemens:2001dx}.  If one
takes small-amplitude sinusoidal wiggles traveling in one direction on
the string and large wiggles (representing the underlying loop in this
case) traveling in the other, and if the difference in wavelengths is
not too large \cite{Siemens:2001dx} one finds that the small wiggles
are damped approximately as an exponential with time constant
proportional to wavelength.  After a fixed time period, wiggles of
different wavelengths would be damped exponentially with exponents
inversely proportional their wavelengths.

Following this idea, we would like to Fourier decompose the
tangent vectors $\ba'$ and $\bb'$ and damp each harmonic $n$ by some
factor $e^{-wk}$, where $w$ is a constant and $k = 2\pi n/L$ is the wave
number.  This corresponds to convolving $\ba'$ and $\bb'$ with a
Lorentzian of width $w$, i.e., $w/(\pi(\sigma^2+w^2))$.

Unfortunately, there is a constraint on the magnitude of the tangent
vectors, $|\ba'|=|\bb'|= 1$.  Smoothing any nontrivial function $\ba'$
or $\bb'$ will decrease its magnitude and violate this constraint.  We
tried simply renormalizing by setting $\ba'\to \ba'/|\ba'|$ after
smoothing, but this works very poorly.  The additional function
$1/|\ba'|$ is in general not very smooth and multiplying by it undoes
most of the desired damping of the high frequency modes.

After trying several techniques to address this problem, we arrived at
the following solution.  Instead of just convolving with a Lorentzian
of width $w$ and then renormalizing, we convolve many times with much
narrower Lorentzians, and renormalize after each one.  We use as our
Lorentzian width 1/20 of the spacing between the points used to
describe the function $\ba'$ or $\bb'$.  (It may seem bit strange to
convolve with something narrower than the point spacing, but since
a Lorentzian falls off slowly, the kernel is still quite wide.)
Distributing the renormalization through the smoothing procedure
yields much better results.  The short modes are in fact damped by
approximately the amount they would have been without renormalization.

We would like to consider loops which have lost some fraction $f$ of
their initial energy to gravitational radiation.  What does this
corresponds to in terms of Lorentzian smoothing?  We consider the
amplitude of the $n = 1$ mode as a proxy for the total length.  This
amplitude is multiplied in the smoothing procedure by $\exp(-2 \pi
c)$, where $c = w/L$ is the ratio of the Lorentzian width to the
string length.  Thus we say that smoothing with coefficient $c$ has
removed a fraction
\be\label{eqn:smoothingfraction}
 f= 1-\exp(-2 \pi c)\,,
\ee
of the initial loop length, and choose $c$ to achieve any desired
fraction of evaporation.

We have studied loops taken from two simulations of size 500 in the
matter era and two of size 1000 in the radiation era.  In each case,
we looked at all loops produced in the second half of the simulation
with the ratio of loop size to horizon size at least 0.01.  For each
loop we repeatedly smooth the loop by convolution, evolve it for at
least one oscillation to check whether smoothing has introduced self-
intersections, smooth the loop again, and so on.  In all there are 7
smoothing steps.  The steps involve amounts of smoothing corresponding
to a total loss since formation of the fractions 1/128, 1/64, 1/32,
1/16, 1/8, 1/4 and 1/2 of the original loop energy.  Since repeated
smoothing with coefficients $c_1, c_2,c_3\ldots$ corresponds to a
single smoothing step with coefficient $c = c_1+c_2+c_3+\cdots$, we
determine the actual smoothing coefficients by solving the equations
\bea
1- e^{2 \pi c_1} &=& 1/128\,,\\
1- e^{2 \pi (c_1+c_2)} &=& 1/64\,,\\
1- e^{2 \pi (c_1+c_2+c_3)}&=& 1/32\,,
\eea
and so on to determine the coefficients $c_1$ through $c_7$.

To simplify the smoothing procedure, which requires repeated fast
Fourier transforms, we work with numbers of frequencies and numbers of
points that are powers of 2.  We start with a function $\ba'$ (and
similarly $\bb'$) that is piecewise constant with non-uniform pieces.
We Fourier transform this function, keeping at step $i$ some number
$N_i$ frequencies (half of which are negative frequencies whose whose
amplitudes are just the complex conjugate of the positive-frequency
amplitudes).  We smooth in Fourier space by multiplying by an
exponential representing the transform of a narrow Lorentzian.  Then
we transform back into $N_i$ uniformly spaced points, adjust the new
tangent vectors to have uniform magnitude, Fourier transform back into
$N_i$ frequencies, and so on, until we have achieved the desired degree
of smoothing for the given step.  We choose
\be
N_i = ( 4096, 4096, 2048, 1024, 512, 256, 128)\,,
\ee
so that frequency $N_i/2$, the highest we keep at each step,
would be reduced by the Lorentzian convolution by about\footnote{The
  actual amount of smoothing is somewhat less than pure Lorentzian
  convolution, because some is undone by renormalization.}  
$10^{-7}$.

The smoothing procedure described above is the best we have found, but
we tried several other procedures (e.g., using a Gaussian instead of a
Lorentzian) and the results are all qualitatively similar.

\section{Smoothing Results}\label{sec:smoothingresults}
\subsection{Smoothing of loops and their tangent vectors}

\begin{figure}
\begin{centering}
\includegraphics[width=.8\linewidth]{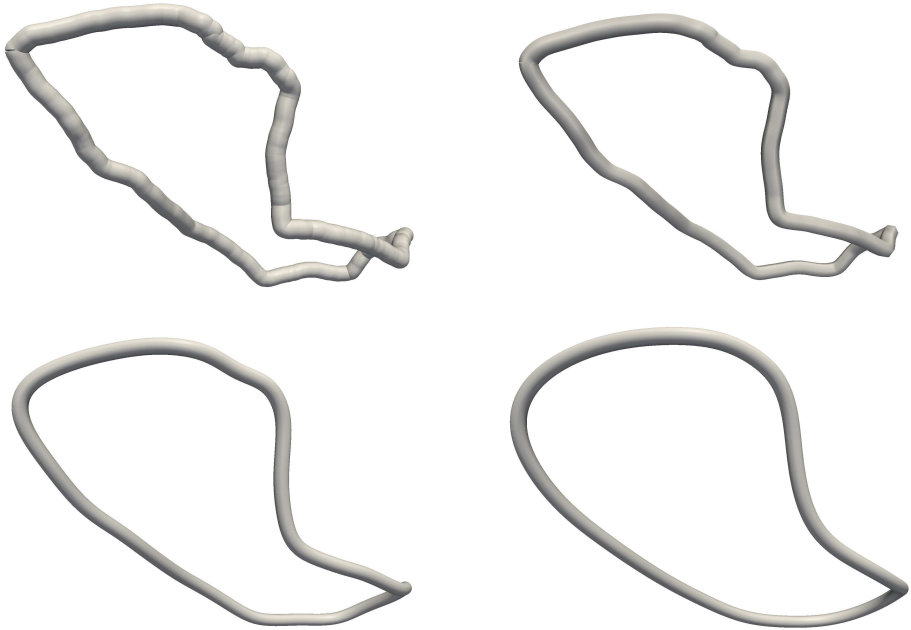}
\end{centering}
\caption{Snapshots for several steps of the smoothing procedure for
  a loop of length 17.9 at time 1000 in the radiation era.}
\label{fig:snapshots}
\end{figure}

We show in Fig.~\ref{fig:snapshots}
some pictures of a typical loop at different stages of the smoothing
process.  We see that indeed our smoothing procedure has removed
small-scale structure from the loop.

\begin{figure}
\begin{centering}
\includegraphics[width=0.8\linewidth]{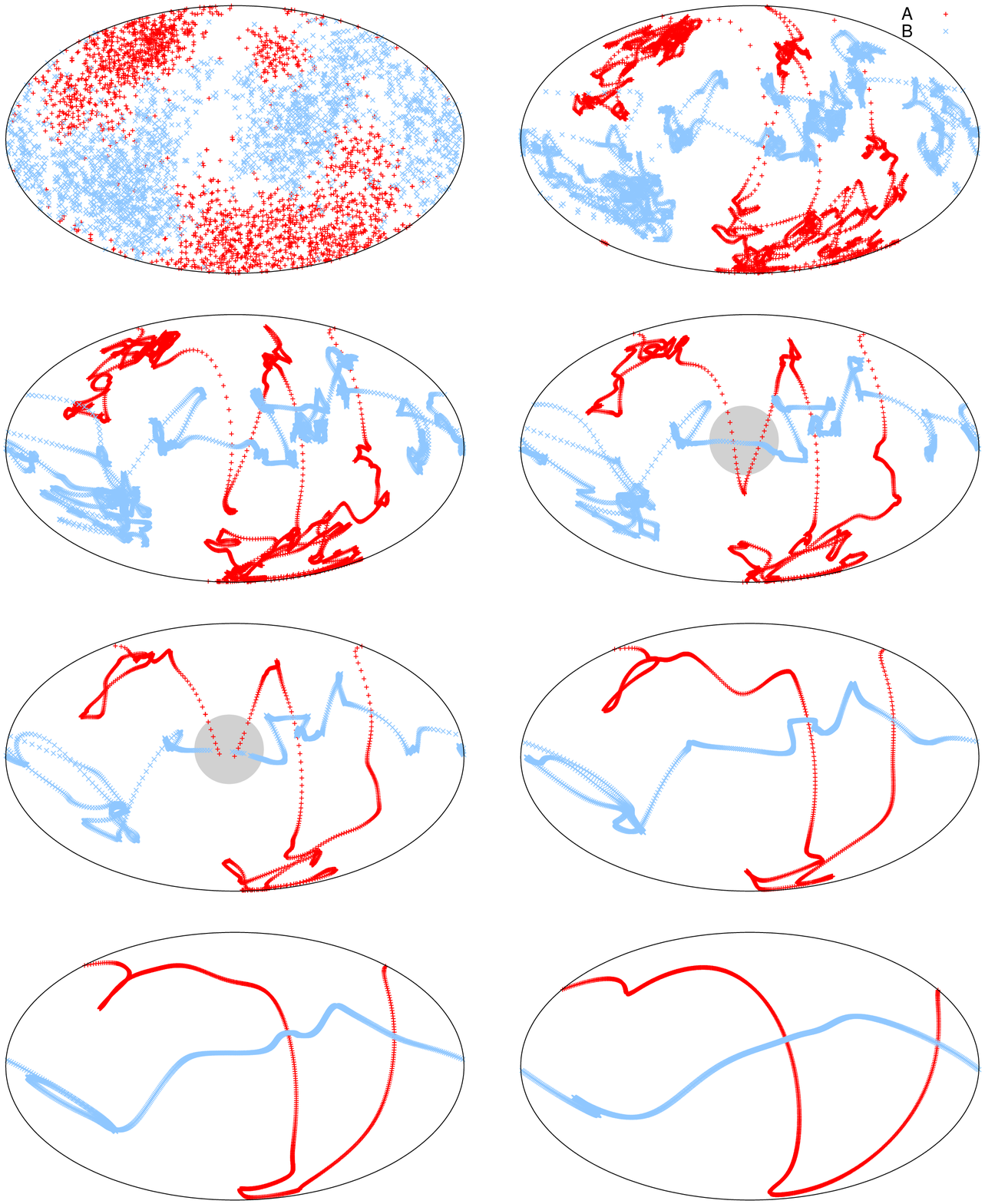}
\end{centering}
\caption{The effect of smoothing on the distribution of the $\ba'$ and
  $\bb'$ in the loop shown in Fig.~\protect\ref{fig:snapshots}. We see
  the production of a small loop from a region of high velocity where
  the ${\bf a' \approx \bf b'}$ after step 3 of our smoothing
  procedure. We mark the region of interest with a shaded disk near
  the center of the figure. The ejection of this loop is seen by the
  disappearance of a series of points of ${\bf a'}$'s as well ${\bf
    b'}$'s. Subsequent smoothing reconnects these points and smooths
  the string again.}
\label{fig:smoothab}
\end{figure}

In Fig.~\ref{fig:smoothab},
we show an example of the change in the paths of $\ba'$ and $\bb'$ on
the Kibble-Turok sphere for a particular loop across the different
stages of smoothing. We can clearly see that the procedure smooths out
structures at increasingly large scales on the figure.  A
fragmentation event can be seen in these figures as a sharp
disappearance of a collection of segments in both the ${\bf
  a'}$'s and the ${\bf b'}$'s. We have chosen a particular loop where
we can see this clearly.  After each fragmentation process, the
remaining loop is not generally in the rest frame, so we boost it back
to its rest frame before we further evolve it.

\subsection{Weighting of simulation data}

When we discuss the fraction of loops with certain properties, we are
interested in the fraction of loops of any given size existing at some
given time.  What the simulation gives us, however, is a set of loops
of different sizes produced in the same interval of time.  As
explained in Ref.~\cite{Blanco-Pillado:2013qja}, each loop should thus
enter into all histograms and averages with weight $x^{\delta}$, where
$x$ is the ratio of the rest energy of the loop in units of $\mu$ to
the horizon distance at the time of formation, and $\delta=1$ for
matter and $3/2$ for radiation.  All results below use this weighting
procedure.

\subsection{Fragmentation}

The first question we ask is whether smoothing causes loops in
non-self-intersecting trajectories to self-intersect and fragment.
The answer is that this is not an important process.  In our
simulations, $90\%$ of the loops survive the smoothing procedure until
half of their energy is eliminated by gravitational radiation with
loss of no more than $10\%$ of their energy into fragmentation. We
have followed this procedure for loops obtained in simulations in the
radiation and matter era and we do not see significant variations on
these numbers for the two cases\footnote{The exact percentages are
$92\%$ for radiation and $87\%$ for matter. If we instead ask of
what is the percentage of loops that lose less than $1\%$ then we
get $79\%$ for radiation and $73\%$ for matter.}.

Looking at movies of the loop evolution after fragmentation we have
identified two major sources of loop fragmentation which are quite
distinct and that pretty much cover all the cases where there is
significant energy lost into loops by the parent loop.

Most of the cases where fragmentation occurs seem to be coming from
fast-moving regions of the parent loop. The loops being produced from
those regions are typically small but often have high Lorentz
boosts. Furthermore there are several cases where we have seen trains
of small loops created from those regions.

The other kind of fragmentation comes from the late time
smoothing of the loop where the smoothing length is already
comparable to the size of the loop. It is not clear how much
one should trust this limiting regime of the smoothing process
so in some sense we may be overestimating the amount of
fragmentation by including these possible energy losses.

\subsection{Number of Cusps}

\begin{figure}
\begin{centering}
\includegraphics[width=.8\linewidth]{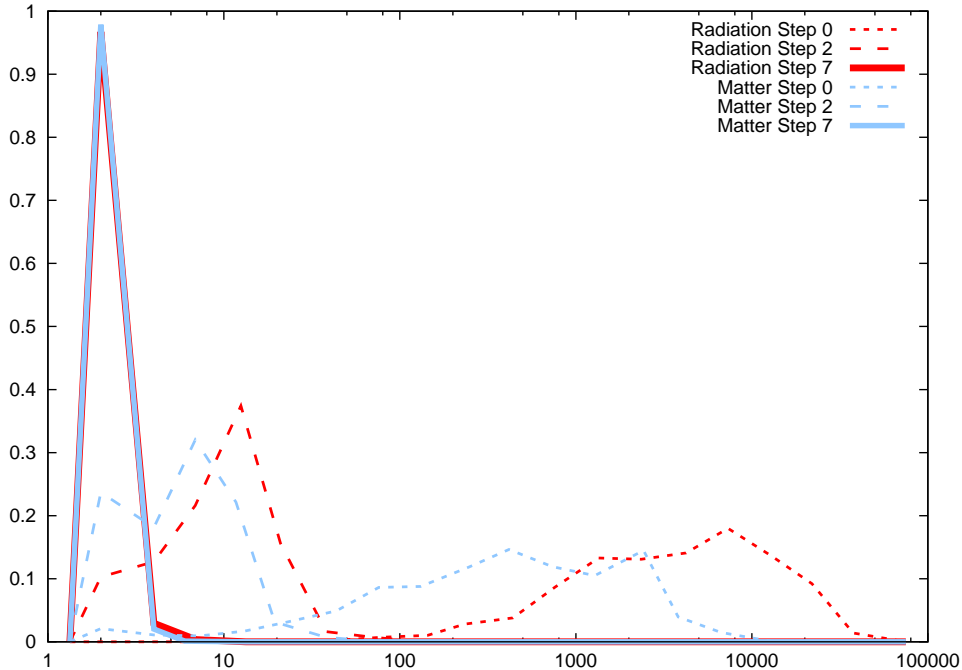}
\end{centering}
\caption{Distribution for the number of cusps before smoothing, after
  two smoothing steps, and when the loop is fully smoothed.  Initially
  there are very many cusps, but these are really just crossings of
  kinks.  By the end there are almost always 2 cusps in both cases.}
\label{fig:cuspnumber}
\end{figure}

As shown in Fig.~\ref{fig:smoothab}, the smoothing procedure yields an
important reduction of the number of crossings between the $\ba'$ and
the $\bb'$ functions on the sphere.  We plot in
Fig.~\ref{fig:cuspnumber} the distribution of the number of cusps for
loops directly from the simulation, after two smoothing steps, and
after the final step of the the smoothing procedure.  We see that the
number of ``cusps'' gets dramatically reduced already in the first
stages of smoothing and quickly settles to 2 cusps per loop either for
radiation or matter.

\begin{figure}
\begin{centering}
\includegraphics[width=.8\linewidth]{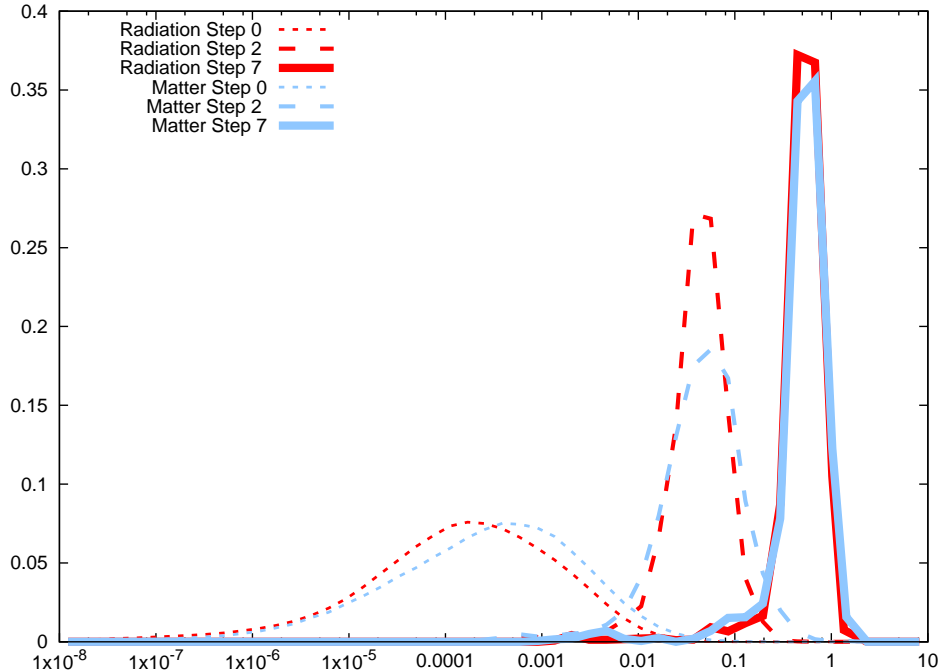}
\end{centering}
\caption{Distribution of the value of the cusp parameter
  $\sqrt{\alpha \beta}$ at the same steps as in
  Fig.~\protect\ref{fig:cuspnumber}.  Initially the ``cusps'' are
  infinitesimal, but by the end they involve most of the loop.}
\label{fig:cuspparameter}
\end{figure}

Furthermore, the amount of energy or length involved in each
crossing increases from a very small amount to a considerable fraction
of the total energy of the loop at the final stage of our smoothing procedure.
In order to quantify this we estimate the value of the length associated with the
second derivative of $\ba$ and $\bb$ at the point of the cusp by calculating
\be
\alpha_{\text cusp} = {{2 \pi}\over {L |{\bf a''}_{\text cusp}|}}\,,
\ee
and similarly for $\beta_{\text cusp}$ in terms of $\bb''$.  We show in
Fig.~\ref{fig:cuspparameter}
the distribution for the values of
$\sqrt{\alpha  \beta}$ at different stages of smoothing.  We use the
product $\alpha\beta$ because this is the area of the world sheet that
is involved in the cusp, and controls, for example, the amount of
gravitational radiation emitted \cite{Damour:2001bk}.

As we mentioned earlier, at the time of loop formation, there are a
very large number of ``cusps'', but these are really crossings of
kinks that have no significant energy as can be seen in
Fig.~(\ref{fig:cuspparameter}). On the other hand, nearly all smoothed
loops have 2 cusps per oscillation, and these cusps are quite
substantial.

\subsection{Angular Momentum}

The angular momentum of a loop is defined by the expression,
\be
{\bf J} = \mu \int_0^L{\left( {\bf x} \times {\bf \dot x}\right) d\sigma}\,,
\ee
which can be written in terms of the the ${\bf a}$ and  ${\bf b}$ functions 
by
\be
{\bf J} = {{\mu}\over {4}} \int{\left( {\bf a} \times \ba' + {\bf b} \times \bb' \right) d\sigma}\,.
\ee

\begin{figure}
\begin{centering}
\includegraphics[width=.8\linewidth]{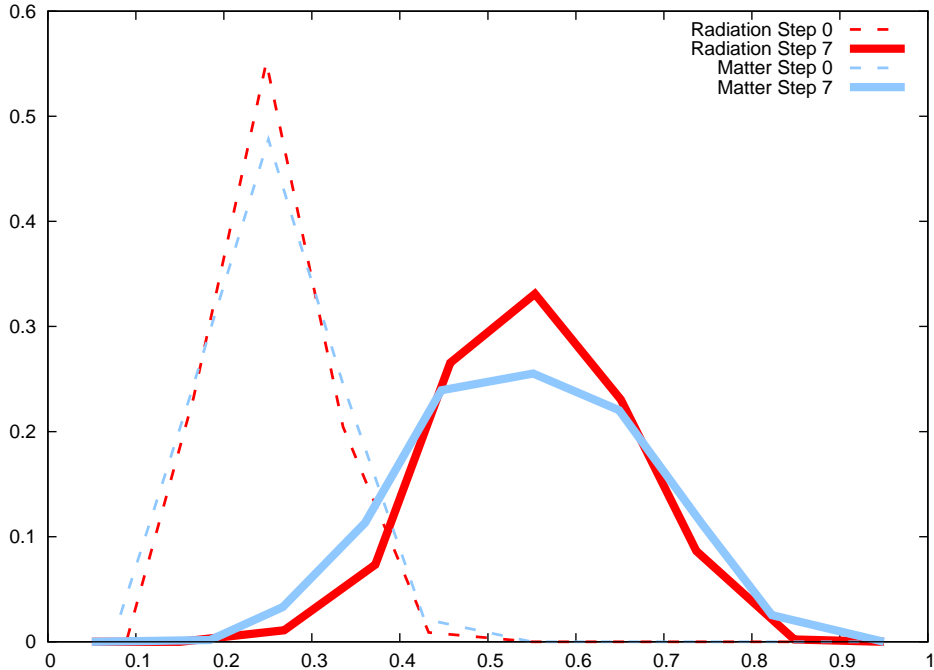}
\end{centering}
\caption{Distribution of the angular momentum of the loops before and
  after smoothing.}
\label{fig:J}
\end{figure}

It is useful to define a dimensionless quantity that compares the 
angular momentum of a loop of certain size $L$ to the maximum angular momentum
for a loop of that energy,\footnote{The state with maximum angular momentum
for a particular mass is the rotating double line, which is the simplest
example of a Regge trajectory \cite{Green:1987sp}.}
\be
{\cal J} =  {{|{\bf J}|}\over {J_{max}}} =  4 \pi{{|{\bf J}|}\over  {\mu L^2 }} \,.
\ee
We show in Fig.~\ref{fig:J} the weighted histogram of the angular
momentum for the distribution of loops in our simulation. The initial
distribution of loops as they come out of the simulation seem to have
a somewhat larger typical angular momentum than the one found earlier
in \cite{Scherrer:1990pj}. This is probably due to the statistical
difference of the loops considered in these earlier studies compared
with our scaling loops.  We see that the distribution moves towards a
larger fraction of the total possible angular momentum after smoothing
has taken place. This does not mean, of course, that angular momentum
increases but only that it grows relative to the total energy on the
loop.  One can understand this by realizing that most of the total
angular momentum of the loop comes from the bulk motion of the loop
and not the small scale structure on it. Smoothing reduces the small
scale structure but does not have a great effect on the angular
momentum.  It would be interesting to see whether a more realistic
model of back reaction changes this picture since gravitational waves
should radiate part of the angular momentum on the loop.

\subsection{Planarity}

Copi and Vachaspati \cite{Copi:2010jw} studied the degree to which
cosmic string loops line in a plane.  Their loops were formed by
fragmentation of single loops with specified harmonics.  Here we do a
similar analysis of loops coming from our simulations and their
smoothed versions.  We consider the tensor
\be
{\cal I }_{ij} = \int dt\,d\sigma \, x_i(\sigma, t) x_j(\sigma, t)\,,
\ee
where the integral is taken over one oscillation and the coordinates
are in the rest frame of the loop.  This tensor (related to the
quadrupole and moment of inertia tensors) gives the extent of the
string world sheet in different directions.  If the motion of the loop
is confined to a plane, ${\cal I }$ will have a zero eigenvalue
in the perpendicular direction.  If the world sheet extends in all
spatial directions equally, then the 3 eigenvalues of ${\cal I }$ will
be equal.  Thus we define the planarity,
\be
 {\cal P}= 1- \left({{\text{ minimum~eigenvalue of~} {\cal I} }\over {\text {average~eigenvalue of~} {\cal I}}}\right)\,,
\ee
which takes on values ranging from 0 for spherical symmetry to 1
for a planar loop.

\begin{figure}
\begin{centering}
\includegraphics[width=.8\linewidth]{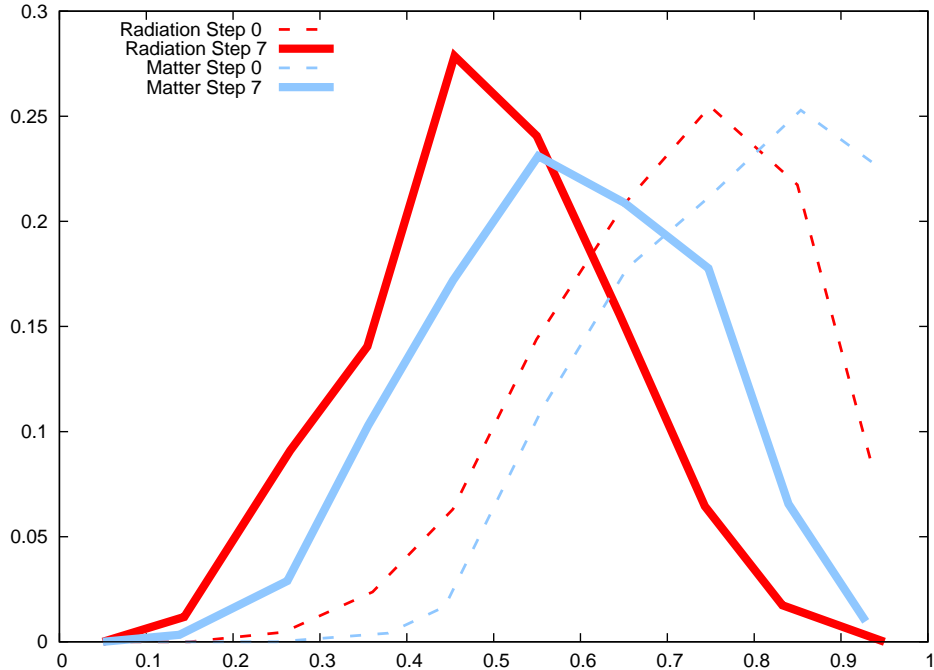}
\end{centering}
\caption{Distribution of the planarity measure ${\cal P}_{\cal I}$ for
  loops at formation and after the last smoothing step.}
\label{fig:planarity}
\end{figure}

We can compute ${\cal I}$ by integrating over $\ba$ and $\bb$
separately.  The results are shown in Fig.~\ref{fig:planarity}.
We find that loops at formation are significantly planar, ${\cal
  P}\sim 0.8$ or 0.9, but smoothing makes them more three-dimensional.

Ref.~\cite{Copi:2010jw} performed the same tensor calculation, but on
$\ba$ and $\bb$ separately, rather than on $\bx$.  They found the two
parts of the string to be substantially linear, with the largest
eigenvalue of order 0.85 of the sum of the eigenvalues.  Adding
together the two mostly linear parts of the string yields a mostly
planar shape, similar to what we found here for loops at formation.
Ref.~\cite{Copi:2010jw} did not attempt to simulate gravitational
smoothing.

\section{Conclusions}\label{sec:conclusion}

We have analyzed the shape of loops and long strings in large cosmic
string network simulations.  Contrary to the usual view that strings
are smooth, we find that long strings consist of generally straight
segments punctuated by large-angle kinks.  We find that the tangent
vectors $\ba'$ and $\bb'$ to the right- and left-moving parts of
strings form hierarchical ``clumps'' on the unit sphere.  We do not
see places where $\ba'$ and $\bb'$ trace out smooth paths.

When a loop forms, it inherits the $\ba'$ and $\bb'$ of the string
from which it formed.  Thus loops at formation do not have smooth
regions of $\ba'$ and $\bb'$ that could cross to form a cusp.
Furthermore, fragmentation of loops nearly completely divides the unit
sphere into regions with only $\ba'$ and those with only $\bb'$.  Only
in unusual cases where these tangent vectors cross into each other's
regions could cusps be possible.

Nevertheless, gravitational back reaction will smooth loops and
produce cusps on loops which formerly had only kinks.  We study this
process with a toy model and find that nearly all loops end up with 2
cusps per oscillation with most of the loop involved in the cusps.

The angular momentum of loops at the time of formation is about one
quarter the maximum possible value, and increases by smoothing into
about half the maximum possible value.  This does not mean that the
angular momentum of the loop is increasing, but rather that the length
of the loop decreases during smoothing and that small-scale
structure that does not contribute much to the overall angular
momentum is the first to be eliminated.

At formation, loops have some degree of planarity, on the order of
0.85 as we described above.  However, smoothing reduces the
degree of planarity so that the smoothed loops are midway between a
planar loop and a loop which has no preferred axes, with a wide
distribution.

\section*{Acknowledgments}

We would like to thank Alex Vilenkin for helpful conversations.  This
work was supported in part by the National Science Foundation under
grant numbers 0903889, 0855447, 1213888 and 1213930. J.J.B.-P. is
supported by IKERBASQUE, the Basque Foundation for Science and the
Spanish Ministry of Science under the FPA2012-34456 grant.

\bibliography{no-slac,paper}

\end{document}